\documentclass{article}
\usepackage{emulateapj}
\usepackage{times}
\usepackage{epsfig}

\newcommand{\kms}{$\mbox{km\,s}^{-1}$}

\begin{document}

\title{Discovery of the optical counterpart and early optical observations
of GRB990712\footnote{Based on observations collected at SAAO, Sutherland;
ESO, Paranal and La Silla (ESO Programs 63.O-0618 and 63-O-0567); 
and AAT, Australia.}}

\author{
K.C. Sahu\altaffilmark{2},
P. Vreeswijk\altaffilmark{3},
G. Bakos\altaffilmark{2,4},
J.W. Menzies\altaffilmark{5},
A. Bragaglia\altaffilmark{6},
F. Frontera\altaffilmark{7},
L. Piro\altaffilmark{8},
M. D. Albrow\altaffilmark{9}, 
I. A. Bond\altaffilmark{10}, 
R. Bower\altaffilmark{11},
J. A. R. Caldwell\altaffilmark{5}, 
A. J. Castro-Tirado$^{12,13}$, 
F. Courbin\altaffilmark{14},
M. Dominik\altaffilmark{15}, 
J.U. Fynbo\altaffilmark{31},
T. Galama\altaffilmark{3,16},
K. Glazebrook\altaffilmark{17},
J. Greenhill\altaffilmark{18}, 
J. Gorosabel\altaffilmark{12},
J. Hearnshaw\altaffilmark{9},
K. Hill\altaffilmark{18},
J. Hjorth\altaffilmark{19},
S. Kane\altaffilmark{18}, 
P. M. Kilmartin\altaffilmark{9}, 
C. Kouveliotou\altaffilmark{20}
R. Martin\altaffilmark{21},   
N. Masetti\altaffilmark{7}
P. Maxted\altaffilmark{22},
D. Minniti\altaffilmark{14},
P. M{\o}ller\altaffilmark{23},
Y. Muraki\altaffilmark{24},
T. Nakamura\altaffilmark{25},
S. Noda\altaffilmark{24},
K. Ohnishi\altaffilmark{26},
E. Palazzi\altaffilmark{7},
J. van Paradijs\altaffilmark{3,27},
E. Pian\altaffilmark{7},
K. R. Pollard\altaffilmark{9}, 
N.J. Rattenbury\altaffilmark{10},
M. Reid\altaffilmark{28},
E. Rol\altaffilmark{3},
T. Saito\altaffilmark{29}, 
P. D. Sackett\altaffilmark{15,17},  
P. Saizar\altaffilmark{30},
C. Tinney\altaffilmark{17},
P. Vermaak\altaffilmark{5},  
R. Watson\altaffilmark{18}, 
A. Williams\altaffilmark{21},
P. Yock\altaffilmark{10}
A. Dar\altaffilmark{32}
}
\altaffiltext{2}{Space Telescope Science Institute, 3700 San Martin Drive, 
Baltimore, MD 21218, USA}
\altaffiltext{3}{Astronomical Institute ``Anton Pannekoek", 
University of Amsterdam, %\& Center for High Energy Astrophysics, 
Kruislaan 403, 1098 SJ Amsterdam, The Netherlands}
\altaffiltext{4}{Konkoly Observatory, PO Box 67, 1525 Budapest, Hungary}
\altaffiltext{5}{South African Astronomical Observatory, P.O. Box 9, 
Observatory 7935, South Africa}
\altaffiltext{6}{Osservatorio Astronomico di Bologna, via Ranzani, 
40127 Bologna, Italy} 
\altaffiltext{7}{Istituto Te.S.R.E., Via Gobetti,
40129 Bologna, Italy}
\altaffiltext{8}{Istituto di Astrofisica Spaziale, CNR, Roma, Italy}
\altaffiltext{9}{Univ. of Canterbury, Dept. of Physics \& Astronomy, 
Private Bag 4800, Christchurch, New Zealand}
\altaffiltext{10}{Faculty of Science, University of Auckland, New Zealand}
\altaffiltext{11}{Univ. of Durham, Dept. of Physics, South Road, Durham,
EEngland DH1 3LE}
\altaffiltext{12}{LAEFF-INTA, P.O. Box 50727, 28080 Madrid, Spain}
\altaffiltext{13}{Instituto de Astrof\'{\i}sica de Andaluc\'{\i}a (IAA-CSIC),
             P.O. Box 03004, 18080 Granada, Spain}
\altaffiltext{14}{Department of Astronomy, P. Universidad Cat\'olica,
Casilla 306, Santiago 22, Chile}
\altaffiltext{15}{Kapteyn Astronomical Institute, Postbus 800, 
9700 AV Groningen, The Netherlands}
\altaffiltext{16}{Palomar Observatory 105-24, Caltech, Pasadena, CA 91125, USA}
\altaffiltext{17}{Anglo-Australian Observatory, PO Box 296, Epping, NSW 2121, 
Australia}
\altaffiltext{18}{Univ. of Tasmania, Physics Dept., G.P.O. 252C, 
Hobart, Tasmania~~7001, Australia}
\altaffiltext{19}{Astronomical Observatory, University of Copenhagen,
Juliane Maries Vej 30, 2100 Copenhagen, Denmark}
\altaffiltext{20}{Universities Space Research Association, 
NASA Marshall Space Flight Center, SD50, Huntsville, AL 35812, USA}
\altaffiltext{21}{Perth Observatory, Walnut Road, Bickley, Perth~~6076, 
Australia}
\altaffiltext{22}{Univ. of Southampton,
Department of Physics and Astronomy, Highfield,
                Southampton SO17 1BJ}

\altaffiltext{23}{European Southern Observatory, Karl-Schwarzschild-Stra\ss e 2
85748, Garching bei M\"unchen, Germany}
\altaffiltext{24}{Solar-Terrestrial Environment Laboratory, Nagoya University, 
               Furocho,Chikusa-ku, Nagoya 464-8601, Japan}
\altaffiltext{25}{Yukawa Institute, Kyoto University, Japan}
\altaffiltext{26}{Nagano National College of Technology, Japan}
\altaffiltext{27}{Physics Department, University of Alabama in Huntsville,
Huntsville, Alabama 35899, USA}
\altaffiltext{28}{Department of Physics, Victoria University of Wellington, New 
Zealand}
\altaffiltext{29}{Tokyo Metropolitan College of Aeronautics, Japan}
\altaffiltext{30}{Lincoln University College, Buenos Aires, Argentina}
 \altaffiltext{31}{Institute of Physics and Astronomy, University of \AA rhus, 
8000 \AA rhus C, Denmark}
\altaffiltext{32}{Department of Physics and Space Research Institute, 
Technion, Haifa 32000, Israel}

\authoremail{ksahu@stsci.edu}

\begin{abstract}
We present the discovery observations of the optical counterpart of the
$\gamma$-ray burster GRB990712 taken 4.16 hours after the outburst and
discuss its light curve observed in the V, R and I bands during the first
$\sim$35 days after the outburst. The observed light curves were fitted
with a power-law decay for the optical transient (OT), plus an
additional component which was treated in two different ways. First, the
additional component was assumed to be an underlying galaxy of constant
brightness.  The resulting slope of the decay is $0.97\pm^{0.05}_{0.02}$ 
and the magnitudes of the underlying galaxy are: 
${\rm V} = 22.3 \pm 0.05$, ${\rm R} = 21.75 \pm 0.05$ and 
${\rm I} = 21.35 \pm 0.05$. Second, the additional component was assumed
to be a galaxy plus an underlying supernova 
with a time-variable brightness identical to that of
GRB980425, appropriately scaled to the 
redshift of GRB990712. The resulting slope of the decay is similar,
but the goodness-of-fit is worse which would imply that either this GRB is not
associated with an underlying supernova or the underlying supernova
is much fainter than the supernova associated with GRB980425.
The galaxy in this case is fainter: ${\rm V} = 22.7 \pm 0.05$, 
${\rm R} = 22.25 \pm 0.05$ and ${\rm I} = 22.15 \pm 0.05$;
and the OT plus the underlying supernova at a given time is brighter.
Measurements of the brightnesses of the OT and the galaxy by 
late-time HST observation and ground-based observations can thus 
assess the presence of an underlying supernova.

\end{abstract}

\keywords{$\gamma$-rays: bursts --- cosmology: observations}

\section{Introduction}

The past two years have witnessed tremendous progress in our understanding of
Gamma-Ray Burst (GRB) phenomena. Thanks to the ability of Beppo-Sax to
provide arcminute sized error boxes for the bursts, the first optical
counterpart was detected in February 1997 for GRB970228 (van Paradijs et
al. 1997; Costa et al. 1997).  The HST observations of this GRB provided the
first clear indication that the GRB is associated with an external galaxy
and is unrelated to its nuclear activity
(Sahu et al. 1997a). Shortly thereafter, the redshift measurement for GRB
970508 proved beyond doubt that GRBs are extragalactic in nature (Metzger et
al. 1997; Djorgovski et al. 1997). More than a dozen GRB optical
counterparts have been detected since then, with redshifts as high as 3.42
in the case of GRB971214 (Kulkarni et al. 1998).

The extensive observations of GRB afterglows in X-ray, optical and radio
wavelengths have been shown to be consistent with cosmological fire-ball
models (e.g. Paczy\'nski \& Rhoads 1993; M\'esz\'aros \& Rees 1997; Wijers
et al. 1997). However, the exact cause of the GRBs has remained elusive, and
progress in determining the actual mechanism causing the bursts has been
slow. The two leading models for GRBs involve the collapse of a massive star
(e.g. Woosley 1997; Paczy\'nski 1998), and the merging of a neutron star
with either another neutron star or a black hole (e.g. Eichler et al. 1989;
Narayan, Paczy\'nski, \& Piran 1992; Sahu et al.  1997b). The ``isotropic
equivalent energy" of some GRBs is as high as 10$^{54}$ ergs, which exceeds
the energy equivalent of the total mass involved in the latter model, and
hence is thought to favor the former.  However, the beaming factor of the
emission of the GRB in different wavebands can be high (in both models), so
the energetics alone may not be conclusive in favoring one of the models
over the other. Since the
lifetime of a very massive star is of the order of a million years or
shorter, the GRBs are expected to be within or close to star-forming
regions in the massive-star collapse model. On the other hand,
since the kick-velocity of a newly formed neutron star is of the order 
of 200 \kms, and since the neutron star, on average, is already about 
10$^8$ years old at the time of the burst, GRBs are generally expected 
to be far from star-forming regions in the merging neutron-star model
(Bloom, Sigurdsson \& Pols 1999).  Thus the location
of the GRB with respect to the star-forming regions in the host galaxy could
be a distinguishing feature which can help in settling the question of the
cause of the GRBs. However, if the presence of dense interstellar material
is prerequisite for the onset of the optical afterglow, then OTs
would be observed {\it only} in star-forming regions regardless of
whether GRBs result from neutron-star mergers or from the collapse of
massive stars. In any case, if some OTs are found far away from
star-forming regions, neutron-star merger model would be favored. 
If all OTs are found in star-forming regions, the situation is less
clear, and depending on the model, one may need to investigate other
aspects to find distinguishing features. For example, one consequence
of the massive-star collapse model is that the GRB should be accompanied by
an underlying supernova, whose brightness variation is distinct from the
power-law decay behavior of the OT and may be detectable. 

%It is worth noting herethat 
There is a  clear bimodality in the observed burst durations: long bursts
have timescales of about 10 to 200 seconds, and short bursts have timescales
of 0.1 to 1 second (Kouveliotou et al. 1995). All the afterglows that have been
discovered so far belong to the subgroup with long bursts (Fishman 1999).
All the OTs for which the host galaxies have been observed
-- although this sample is small -- are found to be in
galaxies whose spectral indicators suggest star-forming activity
(e.g. Fruchter et al. 1999).
So there is an interesting possibility that the 
long-duration GRBs are caused by the collapse of rapidly-rotating massive
stars, and hence they occur in star-forming regions. The short-duration
bursts may be of a different origin, and may be caused by the 
coalescence of two neutron stars. The detection of optical counterparts of short duration bursts,
and their observations would be important to understand this scenario
better.

GRB990712, being at $z=0.434$ (Galama et al. 1999a; Hjorth et al. 2000), 
is one of the closest GRBs
observed so far and hence provides a good opportunity
to determine the possible presence of an underlying supernova,
the location of the GRB within the galaxy, and  the
luminosity of the galaxy in different wavebands. In this paper we present
the optical observations of GRB990712 leading to the discovery of its
optical counterpart,
and subsequent optical imaging observations taken in different filters
during the first $\sim$35 days after the burst.

\section{Observations}

GRB990712 was simultaneously detected on 1999 July 12.69655 with the 
Gamma-Ray Burst
Monitor and the WFC unit 2 onboard the BeppoSAX satellite.  The burst
lasted for about 30 seconds in both the
$\gamma$-ray (40 - 700 keV) and X-ray (2-26 keV) energy ranges and had a 
double-peaked structure.  While its intensity in $\gamma$-rays is
moderate, it exhibited the strongest X-ray prompt emission observed to
date (Heise et al. 1999).

The initial Beppo-SAX position, with an error circle of 5 arcmin radius, was
revised to one with an error radius of only 2 arcmin. Our first observations
were made at this latter position with the SAAO 1m telescope at Sutherland,
South Africa on July 12.87 UT, about 4.16 hours after the burst during
time generally dedicated to PLANET microlensing observations (Albrow et al.
1998).  The optical
image was taken through an R filter, with an integration time of 900 sec.
The detector was a $1024 \times 1024$ SITe CCD with an image scale of 0.309
arcsec per pixel and a total field of view of $5.3 \times 5.3$ square arcmin.
Comparison of the image with the Digitized Sky Survey (DSS) image for the
same field showed the presence of a new source well above the sky
background, that was absent in the DSS image. This new source had a
brightness of R = 19.4 $\pm$ 0.1 mag, which is about 2 magnitudes brighter
than the limiting magnitude of the DSS image, indicating that the new
source was the optical counterpart of the GRB (Bakos et al. 1999a,b). Fig. 1
shows the discovery image, along with the DSS image taken of the same
region. Also marked in the figure are
the reference stars used for photometric calibration of the OT.  A spectrum
of this source obtained a few hours later revealed emission and absorption
lines which were used to derive redshift of 0.434 for this source
(Galama et al. 1999a). Subsequent observations by various groups showed 
the decaying nature of this source further confirming the identification 
of the OT with the GRB.

We continued the observations with a multi-wavelength optical follow-up
campaign using the telescopes at SAAO, ESO and AAO.  A log of all the
observations used in this analysis along with the derived magnitudes of the
OT are listed in Table 1. The measurements include the acquisition images
taken for the spectroscopic observations with the ESO VLT shortly after 
the discovery of the OT (Vreeswijk et al.  1999, in prep.).

\subsection{Astrometry and Photometry of the Optical Counterpart}

An astrometric solution of the field was carried out using 10 reference
stars in the image which are marked in Fig. 1. Their
coordinates, as taken from the USNO 2.0 catalog which uses the Hipparcos
frame of reference, are listed in Table 2. The pixel centroids of the
reference stars were determined using two-dimensional Gaussian fits. These
centroids were combined with the USNO coordinates to determine an
astrometric solution of the field.
The resulting position of the optical counterpart
is RA(2000) = 22$^h$ 31$^m$ 53.$^s 061 \pm 0^s.011$,   Dec(2000)= 
$-73^d \ 24^\prime \ 28.^{\prime\prime}58\, \pm\, 0^{\prime\prime}.05$.
  
Since sky conditions deteriorated due to clouds just after the discovery
observations of the OT, no standard star observations could be taken on that
night. The images were calibrated through observations of the standard stars
F203 and F209 (Menzies et al. 1989) taken on 13th July at SAAO. All the
photometric measurements were carried out using the IRAF aperture 
photometry task 
PHOT. The photometry of a few secondary standards in the field (Fig. 1) 
was carried
out using the standard star observations, and the photometry of all 
GRB observations was then performed via these secondary standards. 
Since only two standard stars were observed, the extinction
correction due to differential airmass between the GRB and the standard star
observations were not applied in the preliminary analysis (Bakos et al.
1999a,b). Instead, the photometric measurements taken with the adjacent 50cm
telescope were used to get the extinction coefficients, which resulted in
a correction of --0.075 mag in V, --0.06 mag in R, and
--0.02 mag in I. (This is consistent with the discrepancy pointed out by Kemp
\& Halpern 1999). Our final magnitudes of the secondary
photometric standards are listed in Table 2, and Table 1 lists the derived
magnitudes of the OT in different bands at different epochs including one
extra measurement reported by Kemp and Halpern (1999).
  
\section{Analysis}
\subsection{The Light Curve in Different Wavebands}

All the photometric measurements listed in Table 1 are also shown in Fig. 2. 
The observed light curves in different bands clearly show a 
continuously-changing slope suggesting a additional component to the
power-law decay of the OT as first noted by Hjorth et al. (1999a). In an 
approach  similar to the one followed by Hjorth et al. (1999b), we have
fitted this additional component in two ways:
(i) with an underlying galaxy of constant brightness, and
(ii) an underlying galaxy plus a supernova similar to GRB980425,
appropriately scaled to the redshift of GRB990712.

First, a power law decline of the form $f(t) \propto t^{-\alpha}$ for the OT
and a constant contribution from the background galaxy was used to fit the
observed light curves simultaneously in different bands, with the same
slope for all the bands. The resultant slope is $0.97$, and the total
$\chi^2$ is 47 for 39 d.o.f.  (Relaxing the condition of the same
slope in different bands does not alter the fit parameters significantly). 
The total $\chi^2$ is, however, fairly flat (between 44 and 52)
for any value of $\alpha$ between 0.95 and 1.02 beyond which the
total $\chi^2$ increases rapidly. Fig. 2
shows the best fit to the light curves.
% where $\alpha$ = 0.97 in all the bands. 
The derived magnitude of the underlying galaxy is
not very sensitive to the exact value of $\alpha$ and 
are found to be
${\rm V} = 22.3 \pm 0.05$, ${\rm R} = 21.75 \pm 0.05$ and 
${\rm I} = 21.35 \pm 0.05$ where the uncertainties represent
the variation of the derived magnitudes as a result of
changing $\alpha$ between 0.95 and 1.02.

One possible consequence of the massive-star model for the GRB is that one 
should observe an underlying supernova in the lightcurve of an OT. 
The physical connection of a GRB with a supernova (SN), first suggested by
the discovery of a peculiar type Ic SN in the error box of GRB\,980425
(Galama et al. 1998; Iwamoto et al. 1998), has been 
strengthened further by recent
observations of other GRBs. Castro-Tirado and Gorosabel (1999)
suggested that the light curve of GRB980326 resembled that of a SN, 
and indeed Bloom et al. (1999) showed that the
late time light curve of GRB\,980326 can be explained by an underlying
SN1998bw type SN at a redshift of around unity. For the afterglow of
GRB\,970228, Reichart (1999) and Galama et al. (1999b) find that a 
power-law
decay plus SN1998bw light curve redshifted to the distance of the burst, $z =
0.695$ (Djorgovski et al. 1999), fits the observed light curve very well. In
light of these new findings, we have also fitted the observed light curve of GRB
990712 assuming the presence of an underlying supernova. 

In order to
determine the contribution of the underlying supernova, we first
calculate the expected V, R, I magnitudes of SN1998bw, placing it at the
redshift of GRB\,990712, $z = 0.434$. This includes
wavelength shifting and time profile stretching (both by a factor of 1+z),
and rescaling the magnitudes for a distance corresponding to $z=0.434$,
assuming %H$_0$ = 70 km s$^{-1}$ Mpc$^{-1}$, and 
$\Omega_0$ = 0.2. To account for the wavelength
shift correctly, we interpolate the redshifted UBVRI broadband flux spectrum
of the SN for each bin in time with a spline fit, obtain the V, R, and I
fluxes at their effective wavelengths, and convert these back to obtain
magnitudes in the observer's frame (Fukugita et al.  1995). The resultant SN
light curves in different bands are shown at the bottom of Fig. 3.
In order to fit the GRB light curve, the SN flux is subtracted from 
the observed GRB flux, and the residual fluxes are assumed to be due 
to the OT and the underlying galaxy.  The procedure outlined earlier 
is then used to fit the light curves. 
The $\chi^2$ in this case is 52 for 42 d.o.f., which is 
clearly higher than the model without the supernova.
The resulting slope of the decay is $0.96$ which is quite similar
but the magnitudes of the underlying galaxy are fainter: 
${\rm V} = 22.7 \pm 0.05$, ${\rm R} = 22.25 \pm 0.05$ and 
${\rm I} = 22.15 \pm 0.05$. The brightness of the OT 
of the GRB at a given time is also slightly lower,
but the combined brightness of the OT and the underlying supernova
(which is the quantity that can be measured if the OT can be
resolved from the galaxy) is higher than the brightness of the OT
in the absence of an underlying supernova. 

It is important to note, however, that there is no evidence  
that the SNe possibly underlying the GRB afterglows
have the same brightness or the same decay
behavior. Our assumption that the underlying supernova is identical to
SN1998bw is dictated by two reasons. First, this is
the only SN associated with a GRB whose light curve has been monitored
extensively. Second, the number of data points in our light curves does not
allow us to vary the characteristics of the underlying SN light curve. 
So the fact that the $\chi^2$ is higher in this case
can be misleading since the true underlying supernova may be different 
from that of SN1998bw, making the resultant $\chi^2$ higher for our
simplified model. This chi-square analysis indicates that
that either GRB990712 is not
associated with an underlying supernova or the underlying supernova
is much fainter than the supernova associated with GRB980425. However, the 
qualitative conclusion that the galaxy is expected to be fainter and the
OT plus the SN is expected to be brighter in the presence of an
underlying supernova is unlikely to change.
Hence late time HST observations, in which the
OT is well resolved so that the brightness of the OT and the galaxy can 
be estimated separately, or late time ground-based observations
when the brightness of the OT is negligible, would greatly help in 
determining the presence of an underlying supernova. 

In both the above scenarios (i.e. with and without an underlying supernova),
the power-law index of the decay $\alpha$ is about $1$, making it one of 
the slowest decline rates of all the OTs observed so far. 
Since $\alpha \le$1 would lead to a divergence of the total energy integrated 
over time, the slope must steepen at later times.
This has been observed for GRB990510 which declined with $\alpha = 0.82$
at early times, later steepening with time (Harrison et al. 1999).
Power-law decays are thought to arise from electrons shocked by the
relativistic expansion of the debris into the ambient medium. In such a case, 
the information on the change of slope can be used to derive information on
the cooling rate of the electrons in the post-shock region
(see, e.g. Wijers et al. 1997; Sari, Piran, \& Narayan 1998;
Livio \& Waxman 1999).

\subsection{Spectral Energy Distribution of the OT}

If the OT emission is due to the synchrotron radiation from the
swept-up electrons in the post-shock region, then its
energy distribution is expected to be of the general form
$f_\nu \propto t^{-\alpha} \nu^\beta$. 
In such a case, one expects a relationship between the power law index 
of the energy distribution of the electrons $p$, the spectral slope
$\beta$, and the decay constant $\alpha$ (Sari et al. 1998). One 
must distinguish two cases: (i) both the peak frequency
$\nu_{\rm m}$ and the cooling frequency $\nu_{\rm c}$ are below the
optical/IR waveband, in which case $p = (4\alpha +2)/3 = 1.97 \pm 0.04$ and
$\beta = -p/2 = -0.99 \pm 0.02$, and (ii) $\nu_{\rm m}$ has passed the
optical/IR waveband, but $\nu_{\rm c}$ has not yet done so, in 
which case $p = (4\alpha +3)/3 = 2.31 \pm 0.04$ and 
$\beta = -(p-1)/2 = -0.66 \pm 0.02$.

We can now compare the theoretically expected spectral slope ($\beta$)
with the observed one.
Although perhaps best determined by the spectroscopic measurements.
our light curve analysis provides magnitudes of the OT 
in different bands, which can be directly used to derive $\beta$. 
We should note here that the brightness of the underlying galaxy
and the possible contamination by an underlying supernova makes the
determination of the magnitudes of the OT less reliable at later times,
and our value of $\alpha$ is most likely dominated by the early part of the
light curve when the OT was bright. Furthermore, $\alpha$
was kept fixed in $all$ the bands for the entire
period of our observations, which implies that 
we cannot detect any spectral {\it evolution} of the OT.
The $synphot/calcphot$ task in IRAF was used to determine
the V$-$R and R$-$I values for a range of $\beta$ values, which were
then compared with the colors derived from the light curve analysis
to determine the spectral slopes.
The V$-$R color implies a spectral slope of $-$1.8 $\pm$ 0.5,
and the R$-$I color implies  a spectral slope of $-$0.7 $\pm$ 0.1.
Thus the slope derived from the V and R bands is closer to the case
(i) mentioned above, and the slope derived from the R and I bands is 
closer to the case (ii). This is roughly consistent with
the theoretical expectations and implies that, in the early part of the 
light curve, $\nu_m$ has passed the V and R and I bands, but 
$\nu_c$ has not passed the I band.

\subsection{The Background Galaxy}

The inferred magnitudes for an underlying galaxy suggest that the galaxy is
relatively bright compared to the OT, and it may be possible to see the host
galaxy directly in the images. The NTT images taken about 8 days after the
burst had the best seeing ($\sim$ 1 arcsec) and the R images indeed show
some hint of a slight extension. To further investigate this, all the images
taken in B,V,R and I bands were co-added, giving rise to a single deep image
of a total integration time of 40 min. From this combined image, we
constructed a model point-spread-function (PSF) from a few isolated bright
stars in the field using IRAF/DAOPHOT.
This model PSF was used to subtract the point source contribution from the
OT.  The PSF-subtracted image shows some residual distribution elongated in
the east-ward direction of the OT over a total of about 2 arcsec, which is
probably the contribution of the underlying galaxy. Thus HST observations
would show the detailed structure of the galaxy (Fruchter et al. in prep). 

The derived magnitudes of the underlying galaxy were used to calculate its
luminosity. The GRB is at a Galactic latitude and longitude of --40 and 315
degrees, respectively. The extinction models of Burstein \& Heiles
(1982) and Schlegel et al. (1998) yield E(B--V) of 
0.015 and 0.033, respectively, for the position of the GRB. This extinction
is small and similar to SN1998bw, which is neglected in our analysis. 
To derive the luminosity of the host galaxy, we use a redshift of 
z=0.434  which, for H$_0$ = 70 \kms and $\Omega$=0.2, corresponds to a
luminosity distance of 2160 Mpc, and a distance modulus of 41.67 mag.
Applying the K-correction for z=0.434, the inferred luminosity of the galaxy
is of the order of $L_\star$ (depending on the presence of an underlying supernova),
where $L_\star$ corresponds to the luminosity of
a typical galaxy at the `knee' of the observed galaxy luminosity function
(see, e.g., Lilly et al. 1999). 
%The host galaxy of GRB990712 is thus
%one of the most intrinsically luminous host galaxies observed so 
%far. 
If we exclude SN1998bw (which  is not only `nearby' ($z = 0.008$), but 
for which the energy released in $\gamma$-rays
is much smaller than any other GRB), then GRB990712 is the 
closest `cosmological' GRB, and the apparent magnitude of its host galaxy is the brightest 
among all the GRB host galaxies observed so far. Compared to the host 
galaxy of GRB990713, which is the next brightest,
the host galaxy of GRB990712 is about 0.25 magnitudes brighter in R 
if there is an underlying supernova, and more than a magnitude brighter if
there is no underlying supernova. 

\acknowledgements{We dedicate this paper to the memory of our colleague and
friend Jan van Paradijs who passed away due to an illness on November 2,
1999, during the final stages of the preparation of this manuscript. 
%Jan was an outstanding scientist who played a key role in the discovery of 
%the first optical counterpart of a $\gamma$-ray burster, GRB970228, and 
%contributed significantly to the recent breakthroughs in this field of research. 
Jan received several awards for his fundamental contributions in this field,
including the 1998 Rossi prize of the American Astronomical Society.  We
will sorely miss his contagious enthusiasm, his brilliant insights, and his
benevolent personality.

This project is partly supported by the Chilean Fondecyt Project 
No.\ 3990024  and the Danish Natural Science Research Council (SNF).
M. Dominik is supported by a Marie Curie Fellowship (ERBFMBICT972457).}

\clearpage 
\begin{figure} %\epsscale{0.6} \plotone{ksahu_fig1a.ps}
%\plotone{ksahu_fig1b.ps} 
\caption{The top panel shows the discovery image of the optical
counterpart (OT) of GRB990712, obtained with SAAO 1m telescope about 4.16 hours
after the burst. The bottom panel shows the Digitized Sky Survey (DSS)
$3 \times 2.8$ square arcmin image of the same field. The
SAAO image shows the optical counterpart (marked `GRB') as a new source
which is about 2 magnitudes brighter than the limiting magnitude of the
DSS image. The reference stars marked 1 and 2 were used as secondary
photometric standards and the reference stars 1 to 10 were used for the
astrometry. North is up and east is to the left.}
\end{figure} 
\clearpage 
\begin{figure} 
\epsscale{0.95} 
\plotone{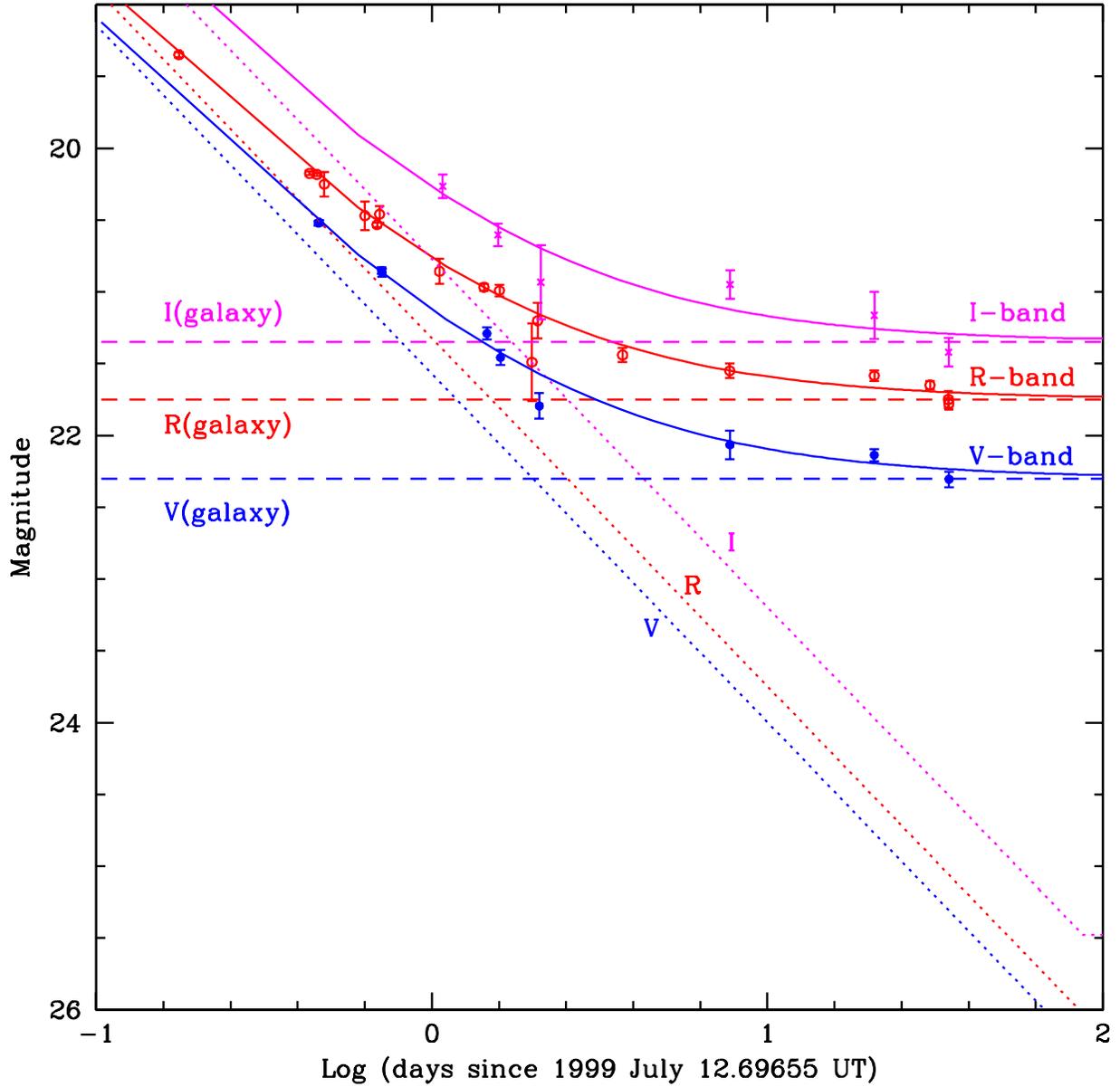}
\caption{The observed light curves of GRB990712 in different
optical bands and the model fits assuming a power-law decay
for the OT and the presence of an underlying galaxy. The I-band measurements
are shown as asterisks, the R-band measurements as open circles, and the
V-band measurements as filled circles. The magnitudes of the underlying galaxy
in different bands are shown as horizontal dashed lines. 
The dotted lines show the decay characteristic of the OT and the solid
curves show the combined OT+galaxy. The derived slope of
the power-law decay is $0.97\pm^{0.05}_{0.02}$ and the magnitudes of the underlying
galaxy are ${\rm V} = 22.3 \pm 0.05$, ${\rm R} = 21.75 \pm 0.05$ and 
${\rm I} = 21.35 \pm 0.05$.}
\end{figure}

\clearpage
\begin{figure}
\epsscale{0.95}
\plotone{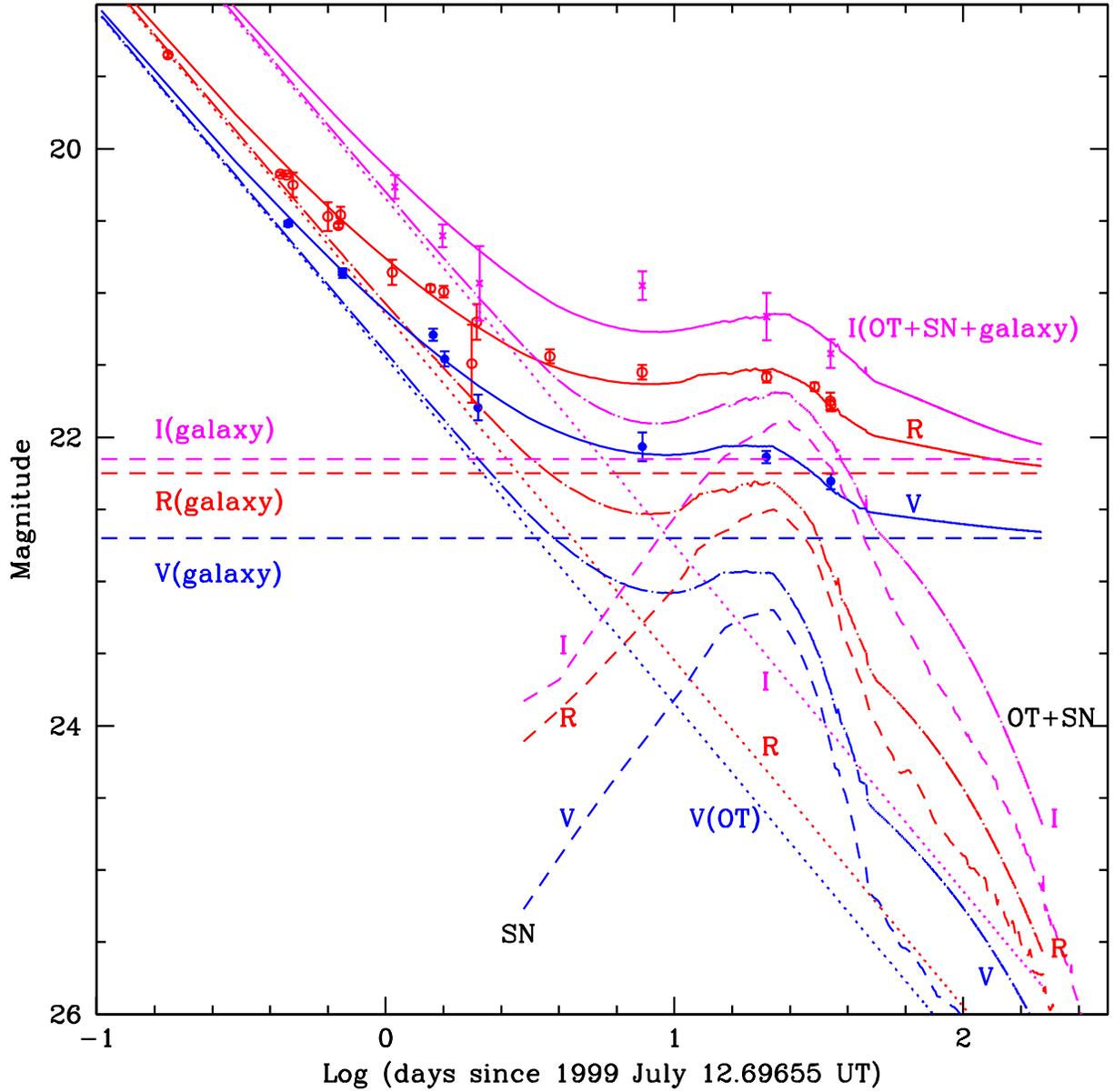}
\caption{The observed light curves of GRB990712 in different
optical bands and the fits to the light curves assuming a power-law decay
for the OT, an underlying galaxy, and an underlying supernova whose
brightness is assumed to be identical to GRB980425 scaled appropriately to
the redshift of GRB990712. The symbols used 
are the same as in Fig. 2. In addition, the light curves of the supernova
(scaled as explained before) in different
bands are shown as dashed lines, the light curves of the
SN+OT are shown as dot-dashed curves, and the light curves of the OT+SN+galaxy
are shown as solid curves. 
The derived slope of the power-law decay is $0.96\pm^{0.05}_{0.02}$
and the magnitudes of the underlying galaxy are
${\rm V} = 22.7 \pm 0.05$, ${\rm R} = 22.25 \pm 0.05$ and 
${\rm I} = 22.15 \pm 0.05$.}

\end{figure}

\begin{deluxetable}{lcll}
\tablecolumns{4}
\tablewidth{0pt}
\tablecaption{Journal of the Observations\label{tab:meas}}
\tablehead{\colhead{Day (UT)} & \colhead{Mag.}
& \colhead{Telescope} }
\startdata
%\\
%\multicolumn{4}{l}{U-band measurements:} &\\
%\\
%\multicolumn{4}{l}{B-band measurements:} &\\
%\\
\multicolumn{4}{l}{R-band measurements:}\\
Jul    12.873&    19.349  $\pm$ 0.019&SAAO 1m& \\
Jul    13.129 &   20.175  $\pm$  0.011&VLT 8m\\
Jul    13.151 &   20.183  $\pm$ 0.009&VLT 8m\\
Jul    13.174 &   20.251  $\pm$ 0.085&ESO 1.5m\\
Jul    13.328 &   20.470  $\pm$ 0.10&MJUO 61cm\\
Jul    13.383 &   20.533  $\pm$ 0.016&VLT 8m\\
Jul    13.395 &   20.460  $\pm$ 0.057&CASLEO 2.15m\\
Jul    13.750 &   20.857  $\pm$ 0.088&SAAO 1m&  \\
Jul    14.127 &   20.968  $\pm$ 0.025&VLT 8m\\
Jul    14.287 &   20.991  $\pm$ 0.040&NTT 3.5m\\
Jul    14.683 &   21.490  $\pm$ 0.27&MJUO 61cm\\
Jul    14.764 &   21.201  $\pm$ 0.123&SAAO 1m&  \\
Jul    16.403 &   21.420  $\pm$ 0.050&NTT 3.5m& \\
Jul    20.421 &   21.550  $\pm$ 0.050& NTT 3.5m\\
Aug    02.533 &   21.584  $\pm$ 0.036&AAT 3.9m\\
Aug    12.232 &   21.650  $\pm$ 0.030&VLT 8m& \\
Aug    16.320 &   21.750  $\pm$ 0.060&CTIO 0.9m$^{\dag}$\\
\tablenotetext{\dag}{Taken from Kemp \& Halpern (1999)}
Aug    16.445 &   21.779  $\pm$ 0.041&AAT 3.9m\\

\multicolumn{4}{l}{V-band measurements:} \\
Jul  13.156&    20.523 $\pm$  0.017     &SAAO 1m\\
Jul  13.158&    20.516 $\pm$    0.015  &SAAO 1m   \\
Jul  13.405&    20.852 $\pm$    0.023  &VLT 8m   \\
Jul  13.406&    20.866  $\pm$   0.028  &VLT 8m   \\
Jul  14.157&    21.290 $\pm$     0.041  &VLT 8m   \\
Jul  14.298&    21.458 $\pm$    0.053  &VLT 8m   \\
Jul  14.787&    21.795  $\pm$   0.089&SAAO 1m&      \\
Jul  20.433&    22.065 $\pm$     0.100& NTT 3.5m     \\
Aug  02.511&    22.136 $\pm$    0.042& AAT 3.9m    \\
Aug  16.471&    22.305  $\pm$   0.054& AAT 3.9m\\
\\
\multicolumn{4}{l}{I-band measurements:} \\
Jul    13.774&    20.266 $\pm$    0.082&SAAO 1m&  \\
Jul    14.271&    20.603 $\pm$    0.078&NTT 3.5m\\
Jul    14.810&    20.933 $\pm$    0.258&SAAO 1m\\
Jul    20.439&    20.950 $\pm$    0.100&NTT 3.5m\\
Aug    02.554&    21.164 $\pm$    0.165&AAT 3.9m\\
Aug    16.414&    21.420 $\pm$    0.100&AAT 3.9m\\
\enddata
\end{deluxetable}

\begin{deluxetable}{lccccccc}
\tablecaption{Positions and Magnitudes of the OT and the reference stars, 
and the magnitudes of the secondary standards. The magnitudes have typical 
uncertainties of 0.01 magnitudes.}
\tablehead{\colhead{No.} & \colhead{RA(2000)} & \colhead{Dec(2000)} &
\colhead{X} & \colhead{Y} & \colhead{V} & \colhead{R} & \colhead{I}}
\startdata
OT & 22:31:53.0614 & -73:24:28.576 &  579.40 & 620.02 \\
1 & 22:31:50.8933 & -73:24:25.280 & 609.98 & 630.95& 17.185 &  16.40 &   15.64\\ 
2 & 22:31:40.1507 & -73:25:09.400 & 758.34 & 487.54& 16.975 &  16.65 &   16.29\\
3 & 22:31:49.7413 & -73:26:09.250 & 624.22 & 294.44& 16.455 &  15.97 &   15.50\\
4 & 22:32:18.6200 & -73:24:09.920 & 224.05 & 680.45& 15.905 &  15.27 &   14.65\\
\\
\multicolumn{8}{l}{Additional reference stars used for astrometry:}\\
 
5 &  22:31:49.3380 & -73:24:53.130 & 631.68 & 537.68 \\
6 &  22:32:02.5307 & -73:25:27.980 & 446.48 & 427.78\\
7 &  22:32:00.6387 & -73:23:20.810 & 474.52 & 839.78\\
8 &  22:31:37.5760 & -73:23:34.160 & 794.42 & 796.29\\
9 &  22:32:21.9453 & -73:24:27.680 & 179.47 & 623.97\\
10 &  22:32:16.1573 & -73:24:55.230 & 259.72 & 535.90
\enddata
\end{deluxetable}

\end{document}